\documentclass[twocolumn,pra,aps,flushbottom,showpacs]{revtex4-1}
\usepackage{xspace,amsmath,amsfonts,amsthm,amssymb,amsbsy,graphicx,color}

\begin{document}

\title{Error-Transparent Evolution: the Ability of Multi-Body Interactions to \\ Bypass Decoherence} 

\author{Os Vy$^{1,2}$, Xiaoting Wang$^1$, and Kurt Jacobs$^{1,2,3}$}

\affiliation{ 
$^1$Department of Physics, University of Massachusetts at Boston, Boston, MA 02125, USA \\
$^2$Advanced Science Institute, RIKEN, Wako-shi 351-0198, Japan \\ 
$^3$Hearne Institute for Theoretical Physics, Louisiana State University, Baton Rouge, LA 70803, USA
} 

\begin{abstract} 
We observe that multi-body interactions, unlike two-body interactions, can implement any unitary operation on an encoded system in such a way that the evolution is uninterrupted by noise that the encoding is designed to protect against. Such ``error-transparent'' evolution is distinct from that usually considered in quantum computing, as the latter is merely correctable. We prove that the minimum body-ness required to protect i) a qubit from a single type of Pauli error, ii) a target qubit from a controller with such errors, iii) a single qubit from all errors, is 3-body, 4-body, and 5-body respectively. We also discuss applications to computing, coherent-feedback control, and quantum metrology.  Finally we evaluate the performance of error-transparent evolution for some examples using numerical simulations. 
\end{abstract}


\maketitle 

\section{Introduction}

Precision control of quantum systems is important in a number of areas. These include potential applications of quantum computing, such as simulating many-body systems~\cite{Abrams97, Wu02, Brown06, Simon11, Barreiro11, Porras12, Casanova12}, quantum metrology~\cite{Wineland92, Wineland94, Huelga97, Shaji07},  and a great variety of experiments that probe quantum behavior~\cite{Hofheinz09, Kubanek09, Barthel10, Leroux2010, Teufel11, Marshall03}. The problem of controlling quantum systems can be divided into two main tasks. The first and simpler task, is that of applying a unitary operator to the system to direct its motion. The second and more general task, is that of modifying the von Neumann entropy of the system. This allows one to combat noise. Because the laws of physics, and thus unitary evolution, are logically reversible~\cite{Plenio01, Maruyama09}, the only way to change the entropy of a system is to transfer this entropy to another quantum system (although sometimes this fact may be obscured by the use of measurement theory). In this paper we concern ourselves only with coherent control (including coherent feedback control) in which no explicit measurements are made. Quantum error-correction (QEC) is an example of a coherent method of combatting noise, and it has the special property that it preserves quantum information stored in the system. A simpler form of control that allows noise reduction is so-called ``coherent feedback'' in which no prior coding is used, but joint unitary operations are applied to the system and an auxiliary~\cite{Lloyd00}. We show here that by using QEC methods, one can design multi-body Hamiltonians whose evolution is unaffected (to first-order) by noise. This can be used, for example, to protect a system from noise in an auxiliary system that is being used to control it. As we show, this can be achieved by replacing the physical auxiliary system with a logical (or ``virtual'') system. We note that ``error-transparent'' evolution is distinct from evolution that is merely correctable --- for example, the transversal evolution that enables fault-tolerant gates~\cite{Gottesman98} is not error-transparent.  

The next section is divided into four parts. In the first we review briefly how error-correcting codes work and define precisely what we mean by ``error-transparency''. We then show how to construct a Hamiltonian that generates error-transparent evolution for a given code. In the second part we discuss the application of error-transparency to coherent feedback control. In the third part we determine  the ``body-ness'' required for a Hamiltonian that is transparent to various kinds of errors. In the last part we discuss briefly under what circumstances it would be advantageous to use many-body interactions that are generated perturbatively. In Section~\ref{appsim} we consider an application of error-transparency to quantum metrology, and present some numerical simulations. We conclude by discussing the origin of the advantage provided by many-body interactions. 
 
\section{Error transparency} 

\subsection{Error-correction codes support Error-transparency}

A traditional quantum error-correcting code (QECC) for the protection of a single qubit works in the following way~\cite{Gottesman09}. We first define a set of operators that form a basis for all single-qubit operations. The basis usually used is the set of Pauli operators, denoted by $X \equiv \sigma_x, Y \equiv \sigma_y$, and $Z \equiv \sigma_z$. The state of a single logical qubit is now \textit{encoded} in a two-dimensional (2D) subspace of a system consisting of $N$ physical qubits. This subspace, called the \textit{code space}, is chosen so that the operation of $X,Y$, or $Z$ on just one of the physical qubits maps the code space to an orthogonal 2D space. The code space is also chosen so that each distinct error either maps the initial space to a distinct orthogonal space, or if any two errors map to the same space, then these two mappings are identical. The spaces to which the errors map the code space are called the \textit{error-spaces}. This construction preserves the encoded information if there is a single error on one of the physical qubits, because the 2D error-space in which the information ends up tells us how the information has been mapped to this space. We can make a measurement to determine which error-space the logical qubit has ended up in, without disturbing the information inside this space. Once we discover where the information is along with what transformation took it there, we then know how it is encoded in this space. This means that we still have access to the information. Note that it is not the state of the $N$-qubit system that is preserved by the error-correcting code, but merely the information that is stored in it. 

During the following discussions we will focus on single operator errors for clarity; however we now take a moment to note that any arbitrary single-qubit error can be written as a superposition of Pauli errors. So if such a superposition afflicts our encoded state, since we have defined our error-spaces in terms of Pauli operators, the logical qubit will become an appropriate superposition of states in more than one error-space. Thus we can still obtain the original information by applying a transformation that maps each error-space back to the code space. To do this we apply an appropriate unitary to all the physical qubits along with an auxiliary system, where this unitary correlates each of the error-spaces with a different basis state for the auxiliary. In other words, this joint unitary applies a different transformation to the physical qubits for each basis state of the auxiliary system. This returns the logical qubit back to the code space with its original information intact, while the unwanted information (entropy) about the superposition that occurred is dumped into the auxiliary system. The ability to correct for a finite set of errors, that forms a basis for all possible errors, allows us to correct any arbitrary single-qubit error.  

If errors happen independently to each physical qubit at a rate $\gamma$, then the probability of a single error occurring on one of the qubits in a time $t$ is $p = n \gamma t$, where $n$ is the number of physical qubits, and this is true so long as $n \gamma t \ll 1$. Usually codes are designed to preserve the encoded information when there is only a single error on any one qubit. This provides an advantage when $p \ll 1$, because in this case the probability that two errors occur is approximately $p^2 \ll p$.  Thus a single-error QECC preserves information under \textit{first-order} effects of independent errors on physical qubits. We now show how such encoding allows logical qubits to have "error-transparent" evolution to the same first-order errors. 

Consider first a Hamiltonian, $H_0$, that performs a transformation \textit{only} on the code-space, so that it transforms the logical qubit. For each error-space we now construct a Hamiltonian that acts on this error-space in a way that is \textit{equivalent} to the action of $H_0$ on the code-space. What we mean by equivalent is the following: the logical qubit is encoded in a specific way on the error-space, and the Hamiltonian that acts on this space performs the same transformation on the logical qubit as $H_0$ performs in the code-space. Let us call the code-space $S_0$ and denote the error (or errors) that take us to the $i^{\mbox{\scriptsize} th}$ error-space by $E_i$, and the Hamiltonian that acts on this error-space by $H_i$. The Hamiltonian $H_i$ will apply a transformation that is \textit{equivalent} to that of $H_0$ if and only if 
\begin{equation} 
    H_i E_i |\psi_0\rangle = E_i H_0 |\psi_0\rangle , \;\;\; \forall |\psi_0\rangle \in S_0 . 
\end{equation} 
Now consider the Hamiltonian, $H = H_0 + \sum_{i=1}^Q H_i$, where $Q$ is the total number of errors in our basis set (three errors for each physical qubit). If a single error occurs, then since the evolution on each of the error-spaces is equivalent to that on the code-space.  The evolution on the logical qubit is not affected in any way by the error regardless of which error-space the information ends up in.  This information will have been transformed in the correct way \textit{despite} the error. This is what we mean by ``error-transparency'' (ET).   

\subsection{Coherent feedback control and ``virtual'' systems}
 
We have seen above that QECC's allow, in theory, ET evolution on a logical system, but do not enable such operations on a physical system. This is interesting from the point of view of coherent feedback control. If we wish to control a physical system, then we cannot use quantum codes to reduce errors in this system. But a quantum controller does not need to be a physical system. If we employ a logical system as a controller --- in which case the controller might be thought of as a ``virtual'' system --- then the target could, in theory, be protected against errors in the controller by using an error-transparent Hamiltonian (ETH). 

\subsection{What degree of ``body-ness'' is needed?} 

The preceding parts discussed the existence of an ETH for logical (encoded) systems; we now focus  on the structure of an ETH. If we write down an arbitrary Hamiltonian for an $N$-body system, it will, in general, have terms that simultaneously connect all $N$ of those bodies. Typically the higher the body-ness of the interaction, the harder it is to come by. So we need to know just what level of body-ness is required for an ETH. 

We can obtain a lower-bound on the body-ness required to realize an ETH for a single qubit, by examining the $\textit{distance}$ requirement of QECC's. To protect against single-qubit errors, the codewords must be chosen so that a single error on one codeword does not produce the same state as any other error on a different codeword. Otherwise it would not be possible to recover the correct initial state from the final state, and the information would be lost. This means, equivalently, that no two errors acting on one codeword can produce another codeword. It therefore requires at least $\textit{three}$ errors to transform one codeword to another; this is described by saying that the code has ``distance three.''  

To produce unitary operations on the code-space, without leaving the code-space, a Hamiltonian must have a matrix element that connects the codewords directly. Since errors on at least three different physical qubits are required to connect any two codewords, a matrix element that connects the codewords must simultaneously change the state of at least three qubits and is therefore a 3-body interaction. 

The above argument does not tell us whether 3-body interactions are sufficient to protect from all single-qubit errors. But we can show that they are sufficient to protect against just \textit{one} of the three Pauli errors (e.g.\ $X$), rather than an arbitrary error. In this case a 3-qubit (classical) code is sufficient. If we only wish to protect against $X$ (bit-flip) errors, then we can use the codewords $|0_{\mbox{\scriptsize L}}\rangle \equiv |000\rangle$ and $|1_{\mbox{\scriptsize L}}\rangle \equiv |111\rangle$ for logical zero and logical one respectively. The code-space is thus $\{|000\rangle,|111\rangle\}$, and the three error-spaces, each corresponding to a bit-flip error on each of the three physical qubits, are $\{|100\rangle,|110\rangle\}$, $\{|010\rangle,|101\rangle\}$, and $\{|001\rangle,|110\rangle\}$. A Hamiltonian that performs a general operation on the code-space is then 
\begin{equation}
  H_0 = a |0_{\mbox{\scriptsize L}}\rangle \langle 0_{\mbox{\scriptsize L}}| + b  |1_{\mbox{\scriptsize L}}\rangle \langle 1_{\mbox{\scriptsize L}}| + c  |1_{\mbox{\scriptsize L}}\rangle \langle 0_{\mbox{\scriptsize L}}| + c^* |0_{\mbox{\scriptsize L}}\rangle \langle 1_{\mbox{\scriptsize L}}| , 
\end{equation}
The corresponding Hamiltonians on each of the error-spaces are obtained by applying each of the errors to $H_0$ (this the general ETH construction). Thus the ETH is~\footnote{If the errors were not self-inverse, then one would have $H = H_0 + \sum_{i=1}^{3} X_i^\dagger H_0 X_i$ instead.} 
\begin{equation}
  H = H_0 + \sum_{i=1}^{3} X_i H_0 X_i . 
\end{equation}

If we want to use the above three-qubit states to control a fourth qubit, the resulting Hamiltonian must now simultaneously change the states of all four systems; it must be 4-body.  Thus an ETH that provides transparency to logical system controlling a "target" system must take the "target" system's body-ness into consideration. 

If we wish to realize evolution that is transparent to all errors on the physical qubits, then the question of the minimal body-ness is more complex. It is clear that a 5-body interaction is sufficient, since a single qubit can be protected from all errors by a 5-qubit code~\cite{Bennett96, Laflamme96}. But it is no longer clear that a 3-body Hamiltonian is sufficient. The code may still be distance three, and the coding states connectable by a 3-body Hamiltonian (this is true of the 7-qubit CCS code). But connecting the two coding states is not all that an ETH must do. It must generate an evolution that is specific to each of the error-spaces. Since each of the error-spaces is defined by the joint states of all the physical qubits, this specificity may require that the action of the Hamiltonian is \textit{conditional} on the states of other qubits. For an action to be conditional on the state of a qubit, that qubit must be involved in the interaction. The total required body-ness is therefore obtained by counting the number of qubits whose state must be changed, as well as those that this change must depend upon. This can also be understood by noting that an action that is conditional on a qubit when viewed in one basis, is instead an active change in the state of that qubit when viewed in another basis. 

We now prove that a 5-body interaction is necessary to realize transparency to all single-qubit errors. We do this by showing that if an $n$-body Hamiltonian exists that is ET for a logical qubit for arbitrary single-qubit errors, then only $n$ qubits are required to correct all single-qubit errors on these same $n$ qubits. But since the smallest single-qubit code has five qubits, $n$ must be no less than five. 

Let us assume that we have a QECC that encodes a single logical qubit in $M$ physical qubits, and that the two logical states are connected by an $n$-body Hamiltonian, $H_0$, with $n<M$. We will refer to the $n$-qubits that $H_0$ acts on as the ``active'' qubits, and the other $M-n$ qubits as the ``passive'' qubits. $H_0$ can perform any operation on the logical qubit, and since it commutes with all errors on the \textit{passive} qubits, it automatically performs the same operation on all error-spaces for these qubits. We therefore only have to worry about the errors on the $n$ active qubits. Since there are three independent errors for each active qubit, there are $3n$ error-spaces for the active qubits. We will label these error-spaces by $j = 1, \ldots, 3n$. Note that for each of the logical states, there is a unique state in each error-space that is the error-state equivalent. We will denote these by $|0j\rangle$ and $|1j\rangle$. Let us now assume that for each error-space there is a Hamiltonian, $H_j$, that performs the correct operation on it (the operational equivalent to $H_0$). Note that the code may be degenerate, so that some of the $H_j$'s may be the same. Note also that each of the Hamiltonians acts as the identity on all the passive qubits. Now we wish to perform error-correction on the active qubits. This means that we need to perform a unitary transformation that maps each of the error-spaces back to the original space. To do this we need to bring up some additional ancilla qubits, because a unitary transformation cannot map from a larger space to a smaller space. Let us denote the states of the ancilla space as $|k\rangle_{\mbox{\scriptsize a}}$. Since the $H_i$'s perform arbitrary operations on each of the error-spaces, we can specialize each of them to an operator $P_j$ that gives $P_j |0j\rangle = |0j\rangle$, and $P_j |1j\rangle = |1j\rangle$. We now use these to form a joint Hamiltonian that correlates each of the error-spaces with an ancilla state. This Hamiltonian is given by 
\begin{equation}
     H = \hbar \lambda \sum_{j=1}^{3n} P_j \otimes \left( |j\rangle_{\mbox{\scriptsize a}}  \langle 0|_{\mbox{\scriptsize a}} +  |0\rangle_{\mbox{\scriptsize a}}  \langle j|_{\mbox{\scriptsize a}}  \right) . 
\end{equation}
Starting the ancilla state as $|0\rangle_{\mbox{\scriptsize a}}$, after a time $\tau = 2\pi/\lambda$ the joint state of the active and ancilla qubits is 
\begin{equation}
   \sum_{j=0}^{3n} \rho_j \otimes  |j\rangle_{\mbox{\scriptsize a}} \langle j|_{\mbox{\scriptsize a}} , 
\end{equation}
where $\rho_j$ is a state confined to the error space $j$ ($j=0$ denotes the code space). Now that the states on each of the orthogonal error spaces are correlated with orthogonal ancilla states, we can apply a unitary operation that maps each error-space back to the code-space. This unitary is
\begin{equation}
  U = U_j \otimes |j\rangle_{\mbox{\scriptsize a}} \langle j|_{\mbox{\scriptsize a}} , 
\end{equation}
where $U_j$ is the error that takes the code-space to error-space $j$ (we are using the fact that the errors are all self-inverse). Since the $E_j$'s are all single-qubit errors on the active qubits, the unitary $U$ does not perform any action on the passive qubits. Since error-correction can be performed for all errors on the active qubits, by only operating on the active qubits, we do not need the passive qubits for error-correction.  Thus $n$ qubits are sufficient for a full error-correcting code. 

As an example, it is quick to show how the 7-qubit CSS code fails to support a 3-body ETH, even though such a Hamiltonian can apply any operation to the code-space. The 3-body logical $X$ operator for this code is $\bar{X} = IIIIXXX$~\cite{Calderbank96, Steane96, Gottesman09}. To obtain a Hamiltonian that acts in the equivalent way on the error space for a $Z$ error, we sandwich $\bar{X}$ between two copies of the operator for this error. This gives, e.g., $IIIIZII \, \bar{X} \, IIIIZII = - \bar{X}$. So if we set $H_0 = \bar{X}$ then one of the error-space Hamiltonians is $H = -\bar{X} = - H_0$. Adding this to $H_0$ cancels both of them, giving zero. Thus there is no 3-body ETH for the 7-qubit CSS code.  

The above proof not only shows that 5-body interactions are required for full error-transparency, but also that if we use an $n$-qubit quantum error-correcting code to create a logical qubit, the error-transparent Hamiltonian for this qubit will require $n$-body interactions. Furthermore, if the logical qubit is being used to control another "target" system, then the ETH will require at least an $(n+1)$-body interaction. 

\subsection{When are perturbative interactions useful?}

Genuine multi-body interactions are harder to come by than 2-body interactions. So let us consider under what conditions one could reduce the effects of noise by generating effective multi-body interactions from 2-body interactions. This can be done perturbatively~\cite{Kempe06, Jacobs09c} using a time-independent perturbation expansion and, in a very similar way, using rapid time-dependent control~\cite{Jacobs07x}. To use the first of these methods we couple the systems for which we want to obtain an all-body interaction (these are the physical qubits of the error-correcting code and any other qubits that the logical qubit might control --- in what follows the ``primary'' systems) to an auxiliary system.  We choose the auxiliary system to have much larger gaps between its energy levels than the primary systems (this means that it also has much faster dynamics than the primary systems), so that the primary systems are a perturbation on the auxiliary Hamiltonian. One then uses time-independent perturbation theory to diagonalize the perturbed Hamiltonian to $k^{\mbox{\scriptsize th}}$-order in the perturbation. If the energy gaps of the auxiliary system have size $\Delta$, and the interaction speed with the auxiliary system has size $\omega$, then the $k^{\mbox{\scriptsize th}}$ order terms in the new effective Hamiltonian contain all $k$-body interactions between the target systems. 

The price one pays for creating a multi-body interaction is a reduction in the rate of the effective dynamics. That is, if one has a 2-body interaction at rate $\omega$, there are two choices. One can use this interaction to perform the task at speed $\omega$ without using an ETH, or one can use it to generate an effective multi-body interaction performing the task error-transparently yet slower. In particular, if one obtains a $k$-body effective interaction by using the $2$-body interaction to perturb a faster single-body dynamics that has rate $\Delta$ (or by using quantum control at rate $\Delta$), then the rate of the effective dynamics is $\omega_{k} = \omega (\omega/\Delta)^{(k-1)}$. Making the ratio $\omega/\Delta$ smaller makes it simpler to obtain the effective interaction with high accuracy.  Now, the effects of decoherence on a transformation are proportional to the decoherence rate, $\gamma$, divided by the speed of the transformation. For a $k$-body interaction generated in the manner above, this speed is $\omega_k$. The probability of an error, $p$, is on the order of $\gamma/\omega_k$. So to see if the error probability is reduced by using an effective multi-body interaction, we need to compare the $p$ that we get when using the 2-body interaction by itself (no error-correction), to that which we get by using an effective $k$-body interaction to implement an error-transparent operation. The former is $p = \gamma/\omega$. When we use an error-transparent Hamiltonian, and thus a quantum error-correcting code, we only have an error in the logical qubit if there are errors in \textit{two} of the physical qubits. So the probability of an error is approximately the square of the probability for a single error to occur, multiplied by the number of physical qubits, $n$. This is $p' = n(\gamma/\omega_k)^2 = n p^2 (\Delta/\omega)^{(2k-2)} $. The use of an effective k-body interaction could thus reduce the effects of decoherence if $n (\gamma/\omega) < (\omega/\Delta)^{(2k-2)}$. Since $\omega/\Delta$ is necessarily less than unity,  effective multi-body interactions will only be useful when $p = \gamma/\omega$ is already rather small.

\section{Simulations}
\label{appsim}

Error-transparent evolution could be useful in any situation in which one is not concerned with the evolution of a particular system but merely with the evolution itself. Examples of this are information processing (computation), coherent feedback control (where the evolution of the controller is merely a means to control a second system), and quantum metrology.

In quantum metrology one wishes to determine the value of a constant (or \textit{parameter}) that appears in the Hamiltonian via the evolution that the Hamiltonian induces in a quantum system~\cite{Kok10}. Such a parameter's value is determines by first preparing some initial state, allowing it to evolve under the Hamiltonian, and inferring the parameter from subsequent measurements.  Such a procedure is clearly protected from noise if the sought after parameter appears in an ETH. Since the physical qubits contain all the information regarding the logical qubit, the parameter can be inferred equally well from a logical system as from a real system.  We note that Caves and collaborators have shown that multi-body interactions can be used to enhance quantum metrology in a completely different manner, by changing the way the accuracy scales with the number of subsystems~\cite{Boixo07, Boixo08}. 
 
We now present some simulations, largely to check that our analysis above is not flawed. Let us say that we have a single qubit that evolves under the Hamiltonian $H = \hbar \omega Z= \hbar \omega ( |0\rangle \langle 0| - |1\rangle \langle 1| ) $, and we want to protect this from $X$ errors. To to this we use a our previously discussed three-qubit encoding into logical states: $|000\rangle$ and $|111\rangle$. The equivalent evolution to this logical code-space becomes $H_0 = \hbar \omega ( |000\rangle \langle 000| - |111\rangle \langle 111| )$. Now to turn this into an ETH, we add a new term with each error appropriately applied to $H_0$:
\begin{equation}
  H = H_0 + IIX H_0 IIX + IXI H_0 IXI + XII H_0 XII .
\end{equation}
Since the logical qubit is transparent only to a single error, its effective error rate will be lower only for times short compared to the inverse error rate. Measuring time in units of the oscillation period $\tau = \pi/\omega$, we choose the single-qubit error rate to be $\gamma = \alpha(1/\tau)$ where $\alpha \ll 1$. The error probability for a physical qubit during this time is $p = \gamma\tau = \alpha$. The total probability of an error occurring on any of the three physical qubits is therefore approximately $p_{\mbox{\scriptsize tot}} = 3p$. Since, in our error-transparent framework, the logical qubit only experiences an error when there are at least \textit{two} errors on the physical qubits; the effective error rate for the logical qubit is lower. The probability that there are two errors in time $\tau$ on any of the three qubits is $p_{\mbox{\scriptsize tot}}^2$, but we note that two errors on a single qubit cancel each other out. Thus there are just three ways to obtain two errors and so the approximate error probability for the logical qubit in time $\tau$ is $p_{\mbox{\scriptsize L}} = 3 p^2$. The effective error rate for the logical qubit is then $\gamma_{\mbox{\scriptsize L}} =  3 p^2/\tau = 3\alpha^2/\tau = 3\alpha\gamma$. So the code, along with the ETH, reduces the effective error rate by a factor of approximately $3\alpha$; this will guide us in comparing the simulations.    

\begin{figure}[t]
\leavevmode\includegraphics[width=1\hsize]{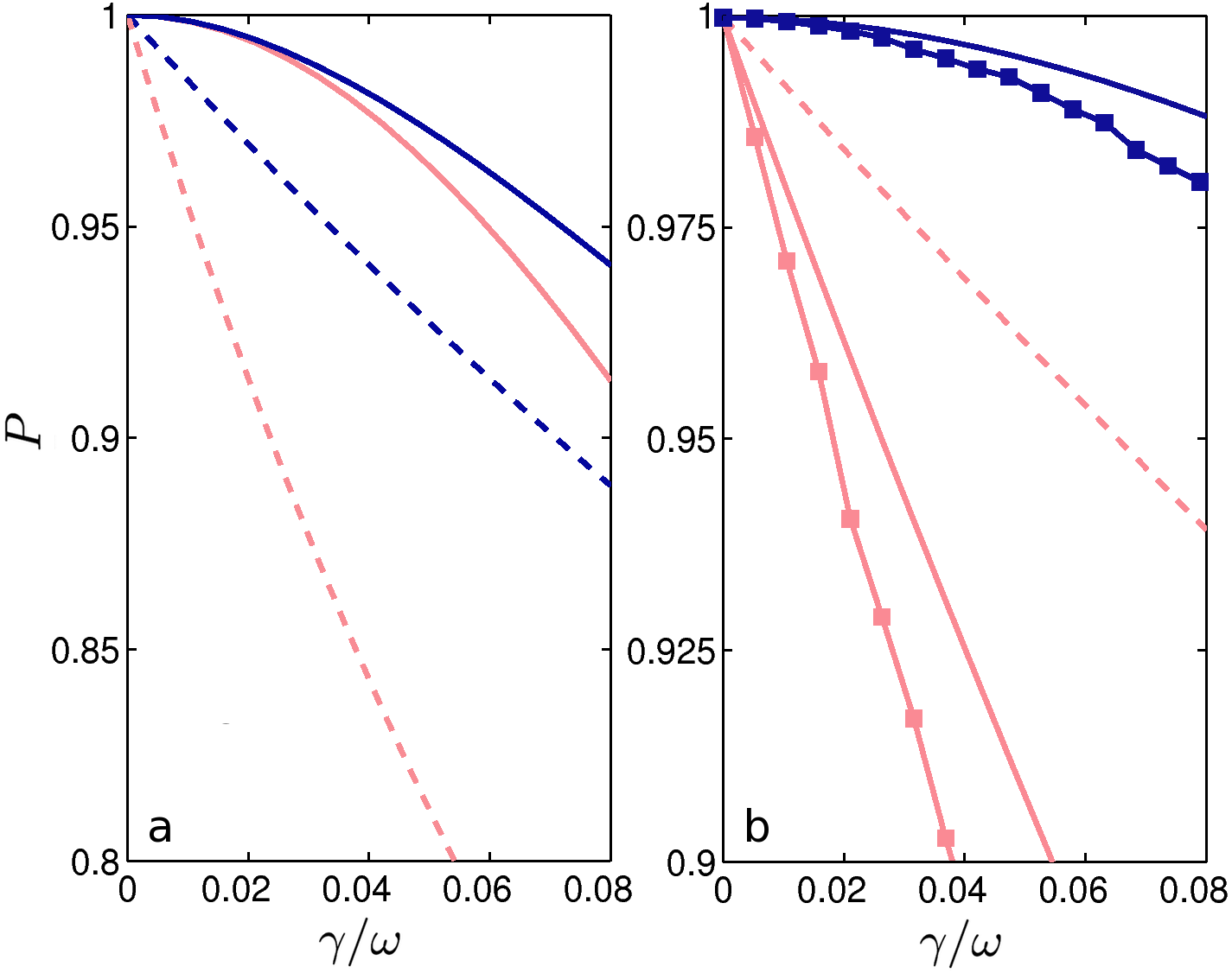}
\caption{(Color online) Here we show the increased fidelity that can be achieved using error-transparent Hamiltonians (ETH's). The probability of finishing in the desired state is shown verses the ratio of the error-rate to the swap rate ($\gamma/\omega$), for two scenarios: (a) Four scenarios of a single and logical qubit evolving under $Z=\sigma_z$ with frequency $\omega$ subjected to $X=\sigma_x$ noise. Dark dashed line: single qubit with no error-correction or transparency; light solid line: a single qubit with an effective second-order error-rate realized by the ETH; light dashed line: three qubit logical state with no transparency; dark solid line: three-qubit logical state with a 3-body ETH (b) A single qubit being controlled by various logical qubit controllers. Light solids lines: control via a logical qubit encoded with 5-qubit and 7-qubit codes (the latter uses a Monte-Carlo simulation and so the squares give approximate error-bars); dark solid lines: the equivalent logical controllers with ETH's; light dashed line: control via a single, unprotected, controller qubit.} 
\label{fig1} 
\end{figure} 

We perform simulations to compare this evolution in four scenarios: i) a single physical qubit evolving under $H$ with $X$ errors at rate $\gamma$; ii) the same single qubit with $X$ errors but at rate $0.03\gamma$; iii) the 3-qubit logical state evolving under X noise on each of its three physical qubits; and iv) the same logical state with $X$ errors but this time evolving under the ETH. For each of these scenarios we start the qubit in state $|1\rangle + |0\rangle$ (and the equivalent logical state for the last two scenarios) and evolve them for one cycle; then we check the probability that they go back to their initial states. In Fig.~\ref{fig1}a we plot this probability as a function of the error-rate $\gamma$ (always divided by the constant $\omega$ to make it dimensionless). We see from these plots that the ETH, the best preforming out of the four, indeed suppresses the errors in its evolution and transforms them, as expected, from first to second-order errors. Further, we see that for small $\gamma\tau$ the logical qubit evolves in the same way as a physical qubit with the reduced error-rate given by $3\alpha\gamma$. We also show that the logical qubit evolves under $H_0$ rather than the ETH, its error rate is faster than each physical qubit, as we would expect. 

As our second example, we consider coherent feedback control of a single ``target'' qubit by a logical qubit encoded using the 5-qubit stabilizer code~\cite{Bennett96, Laflamme96} and the 7-qubit CSS code~\cite{Calderbank96, Steane96, Gottesman09}; the added target qubit giving us 6 and 8-body dynamics respectively. In all cases we start the target in the ground state coupled to a excited logical state, and the Hamiltonian's goal is to swap these during a set time in a thermalization environment.  Our target qubit will be subjected to both excitation and damping noises with constant rates of $10^-4$ and $20^-4$ respectively.  The comparatively larger auxiliary damping rate, $\gamma$, will range from 0 to 0.1 and will be applied to each of the qubits composing the logical controller. Our success, like in the previous example, will be measured by the probability of the target qubit finishing in the excited state, and we will plot this against the ratio of the auxiliary noise to the swap rate $\omega$. This is done for five different scenarios.  The first two will be preforming the swaps under the non-ET coupling. The next two will be the same swaps with full auxiliary ETH's (including protection from Z and Y errors now).  Our final scenario will be the same swap with just a single "logical" qubit.  Meaning that we suspect the ETH cases will out-perform their less protected counterparts, yet we also need them to out-preform using just one noisy qubit, or we will find little practical value of adding those ET five or seven auxiliary qubits. In this example, the physical qubit could represent the first two states of a nano-mechanical oscillator that we wish to prepare in the one-phonon Fock-state.  

We display the results in Fig.~\ref{fig1}b. The results show us that indeed the control achieved on the target qubit is greatly improved using the ETH's. Further, as expected, the 5-qubit code performs better than the 7-qubit code, since in the latter's effective physical error-rate is greater by a factor of $7/5$. We also see, as expected, that the logical qubits perform much worse that a single-qubit controller when they do not have an ETH.  Yet on the other hand, we thankfully see that both of the ETH cases out-preform using just a single, noisy, controller qubit. 

\section{Conclusion}

In this paper we have shown that multi-body interactions can generate evolution that is uninterrupted by single-qubit errors. It is worth noting that this ability can be viewed as resulting from the fact that the body-ness of an error-transparent Hamiltonian is greater than those generating the errors. The error-correcting codes we have considered here work precisely because the errors on different subsystems are assumed to be independent. We note that if the subsystems that make up the bath were to couple to the physical qubits via $k$-body interactions, then they would likely induce errors that were correlated across $k-1$ physical subsystems. In this case, $k$-body interactions would not be able to reduce the effects of the errors. One can therefore summarize the power of multi-body interactions to implement error-transparent operations in the following way: error-transparent operations can be realized in a multi-body system if the bath interacts with the system via $k$-body interactions, and the bodies of the system interact with each-other via $m$-body interactions, with $m>2k$. Of course, this characterization is blurred somewhat by the fact that, given appropriate timescale separations, effective multi-body interactions can be used to obtain a degree of error-transparency. 

\section*{Acknowledgments} 

KJ and XW are partially supported by the NSF under Project No.\ PHY-0902906, and by IARPA (see~\footnote{Partially supported by the Intelligence Advanced Research Projects Activity (IARPA) via Department of Interior National Business Center contract number D11PC20168. The U.S. Government is authorized to reproduce and distribute reprints for Governmental purposes notwithstanding any copyright annotation thereon. Disclaimer: The views and conclusions contained herein are those of the authors and should not be interpreted as necessarily representing the official policies or endorsements, either expressed or implied, of IARPA, DoI/NBC, or the U.S. Government.}). KJ and OV are partially supported by the NSF under Project No.\ PHY-1005571, and by the ARO MURI grant W911NF-11-1-0268.  


%

\end{document}